\documentclass[pra,twocolumn,showpacs,superscriptaddress,10pt]{revtex4}
\usepackage{pifont}
\usepackage{amsmath}
\usepackage{amssymb}
\usepackage{amsfonts}
\usepackage{times,txfonts}
\usepackage{bbold}
\usepackage{hyperref}
\hypersetup{
    colorlinks=true,
    citecolor=blue,
    linkcolor=blue,
    urlcolor=blue
}
\usepackage{color}
\usepackage{bm}
\usepackage{graphicx}
\usepackage{float}
\usepackage{epsfig}
\usepackage{verbatim}
\usepackage{epstopdf}

\newcommand{\ket}[1]{\left\vert#1\right\rangle}
\newcommand{\bra}[1]{\left\langle#1\right\vert}
\newcommand{\braket}[2]{\langle{#1}|{#2}\rangle}

\newcommand{\inner}[2]{\left\langle#1\kern-\nulldelimiterspace\left|#2\kern-\nulldelimiterspace\right.\right\rangle}
\newcommand{\tr}{\mathrm{Tr}}
\newcommand{\im}{\mathrm{Im}}
\newcommand{\re}{\mathrm{Re}}

\def\half{\frac{1}{2}}

\def\mean#1{{\langle #1 \rangle}}
\def\etal{{\it et al. }}
\def\ie{{\it i.e.}}
\def\eg{{\it e.g.}}

\begin{document}

\title{Inseparability Criterion Using Higher-Order Schr\"odinger-Robertson Uncertainty Relation 
}

\author{Chang-Woo Lee}
\address{Department of Physics, Texas A\&M University at Qatar, PO Box 23874, Doha, Qatar}
\author{Junghee Ryu}
\address{Institute of Theoretical Physics and Astrophysics, University of Gda\'{n}sk, 80-952 Gda\'{n}sk, Poland}
\author{Jeongho Bang}
\address{Center for Macroscopic Quantum Control \& Department of Physics and Astronomy, Seoul National University, Seoul, 151-747, Korea}
\address{Department of Physics, Hanyang University, Seoul 133-791, Korea}
\address{Institute of Theoretical Physics and Astrophysics, University of Gda\'{n}sk, 80-952 Gda\'{n}sk, Poland}
\author{Hyunchul Nha}
\address{Department of Physics, Texas A\&M University at Qatar, PO Box 23874, Doha, Qatar}

\date{\today}

\begin{abstract}
We formulate an inseparability criterion based on the recently derived
generalized Schr\"odinger-Robertson uncertainty relation (SRUR) [Ivan {\it et al.} J. Phys. A :Math. Theor. {\bf 45}, 195305 (2012)]
together with the negativity of partial transpose (PT). 
This generalized SRUR systematically deals with two orthogonal quadrature amplitudes to higher-orders, so is relevant to characterize non-Gaussian quantum statistics.
We first present a method that relies on the single-mode marginal distribution of two-mode fields under PT followed by beam-splitting operation.
We then extend the SRUR to two-mode cases and develop a full two-mode version of inseparability criterion. 
We find that our formulation can be useful to detect entanglement of non-Gaussian states even when, e.g., the entropic criterion that also involves higher-order moments fails. 
\end{abstract}


\maketitle

\section{Introduction}
Uncertainty relation (UR) has played a fundamentally crucial role 
in characterizing quantum mechanics ever since its birth.
UR can also be employed as a pivotal tool in quantum information tasks, e.g.
entanglement detection.
In particular, tests adopting matrices of moments (MMs) together with negativity of partial transposition (NPT) have been broadly used 
\cite{Simon00,Duan00,Mancini02, 
Raymer03,Agarwal05,Shchukin0511,Hillery06,Nha06
}.
The principle behind such tests is that the
negativity of the partially transposed density matrix of an entangled state also induces the negativity of its MM.
As a matter of fact, provided that an infinite hierarchy is allowed, the converse is also true \cite{Shchukin0511,Miranowicz09},
that is, 
if the MM of a partially transposed density matrix is negative then 
the state is NPT entangled.
Criteria utilizing MM and NPT are particularly useful for 
continuous-variable (CV) states. For the discrete variable case, numerous measures and detection schemes for entanglement are well developed,
which, however, is more challenging for the CV case except very special ones.
This is because the Hilbert space of CV is infinite dimensional and thereby can have a more involved structure.

In this respect,
Simon essentially derived a second-order MM for an inseparability criterion \cite{Simon00},
which is a necessary and sufficient condition for Gaussian states.
Duan \etal also proposed a complete criterion based on MM for Gaussian states \cite{Duan00} 
although it does not seem to be directly connected to NPT.
An entangled Gaussian state does not need any higher-order MM to detect its inseparability.
Since then, a number of other criteria have also been developed 
in order to address also non-Gaussian entangled states \cite{Mancini02,
Raymer03,Agarwal05,Shchukin0511,Hillery06,Nha06,
Miranowicz10,Walborn09,Nha12}.
Of these criteria, Shchukin and Vogel (SV)'s
is recognized as a unified one that includes in its hierarchy
all the MM inseparability criteria \cite{Shchukin0511}.
Later,
the above MM inseparability criteria has been analyzed in terms of
MM of normally ordered operators \cite{Miranowicz10}, 
which was also developed by SV aiming at a nonclassicality criterion
\cite{Shchukin0501}.

In this article,
in the same spirit of the above approaches---namely, 
based on MM and NPT---we 
also propose a unified inseparability criterion, especially
focusing on the fourth- and even higher-order moments.
Our study is based on a generalized Schr\"odinger-Robertson uncertainty relation (SRUR)
recently derived by Ivan \etal \cite{Ivan12a
}.
This naturally includes the original SRUR in its hierarchy and 
extends it to arbitrary (higher) orders for a single-mode system.
We first develop an inseparability criterion by applying this single-mode SRUR to a marginal distribution of 
a partially transposed bipartite system. We next extend the single-mode SRUR to a two-mode system and subsequently derive a full two-mode SRUR inseparability criterion.
We show that our criterion is actually equivalent to SV's, 
hence is a unified criterion and can detect any NPT entanglement. 
On the other hand, our formulation explicitly addresses two orthogonal quadrature amplitudes rather than the creation and the annihilation operators unlike other criteria \cite{Shchukin0511,Miranowicz10}. We also illustrate that our criterion can be more powerful in detecting certain classes of non-Gaussian entangled states than the entropy-based criterion \cite{Walborn09} although the latter also addresses higher-order moments in a specific form.

Our approach as well as other ones based on MM is experimentally feasible and dose not need a full tomography. 
Along with theoretical schemes \cite{Shchukin0501,Shchukin0510},
experimental technology is being continuously developed.
For instance, refer to a recent breakthrough about detecting higher-order moments of correlation \cite{Menzel10,Eichler11}.

\section{Generalized Schr\"odinger-Robertson uncertainty relation for single-mode case}
\label{sec:srur1}
First,
we briefly review the generalized Schr\"odinger-Robertson UR (SRUR)
recently derived by Ivan \etal \cite{Ivan12a
}.
This generalized version involves moments higher than the second and naturally include the second-order ones in its hierarchy.
We begin by noting that 
every uncertainty relation associated with moments relies on the positivity (precisely, positive semi-definiteness) of a density matrix $\rho$.
For an arbitrary linear combination of quantum operators
$\hat F = c_0 \hat 1 + c_1 \hat f_1 + c_2 \hat f_2+ \cdots+ c_n \hat f_n$
($c_i$'s are c-numbers, $\hat 1$ is the identity operator, and $\hat f_i$'s need not necessarily be hermitian),
the mean value of the product of its adjoint and itself must be non-negative, \ie,
\begin{equation}
\mean{\hat F^\dag \hat F} 
= \tr \left(\rho \hat F^\dag \hat F \right)
= {\bf c}^\dag M_{\hat {\bf f}} (\rho) {\bf c} \ge 0,
\end{equation}
where ${\bf c} = \left( c_0,~c_1,~c_2,~\cdots,~c_n \right)^T$,
\begin{equation}
M_{\hat {\bf f}} (\rho) = 
\left[ \begin{array}{cccc} 
\mean{\hat f_0^\dag \hat f_0} & \mean{\hat f_0^\dag \hat f_1} & \cdots & \mean{\hat f_0^\dag \hat f_n}\\
\mean{\hat f_1^\dag \hat f_0} & \mean{\hat f_1^\dag \hat f_1} & \cdots & \mean{\hat f_1^\dag \hat f_n}\\
\vdots & \vdots & \ddots & \vdots \\
\mean{\hat f_n^\dag \hat f_0} & \mean{\hat f_n^\dag \hat f_1} & \cdots & \mean{\hat f_n^\dag \hat f_n}
\end{array} \right],
\label{eq:mf}
\end{equation}
and $\hat {\bf f} = (\hat f_0,~\hat f_1,~\hat f_2,~\cdots,~\hat f_n )$ with $\hat f_0 = \hat 1$.
The inequality (1) must be satisfied for arbitrary $c_i$'s, which implies that 
the matrix of moments, $M_{\hat {\bf f}}$, should be positive semi-definite, 
\ie, 
\begin{equation}
M_{\hat {\bf f}} (\rho) \ge 0.
\label{eq:urmm}
\end{equation}
Note that the above inequality employing $\hat {\bf f} = (\hat 1,~\hat f_1,~\cdots,~\hat f_n)$
can be made equivalent to $M_{\hat {\bf f}'}  (\rho)\ge 0$ where
$\hat {\bf f}' \equiv(\Delta \hat f_1,~\cdots,~\Delta\hat f_n)$ refers to a variance operator $\Delta \hat O \equiv \hat O - \mean{\hat O}$
\cite{Miranowicz10,Ivan12a
}. Henceforth we will consider only the latter case to our aim.

In fact, the second-order SRUR of any hermitian operators $\hat A$ and $\hat B$ is equivalent to 
$M_{\hat {\bf f}} \ge 0$ with $\hat {\bf f} = \left( \Delta \hat A, \Delta \hat B \right)$,
more specifically,
\begin{equation}
\left| \begin{array}{cc} 
\mean{\Delta\hat A^\dag \Delta\hat A} & \mean{\Delta\hat A^\dag \Delta\hat B} \\
\mean{\Delta\hat B^\dag \Delta\hat A} & \mean{\Delta\hat B^\dag \Delta\hat B} 
\end{array} \right|  
\ge 0.
\end{equation}
This inequality 
\begin{equation}
\mean{\Delta\hat A^\dag \Delta\hat A} \mean{\Delta\hat B^\dag \Delta\hat B} \ge
|\mean{\Delta\hat A^\dag \Delta\hat B}|^2
\end{equation}
is tighter than the Heisenberg UR (HUR)
\begin{equation}
\mean{\Delta\hat A^\dag \Delta\hat A} \mean{\Delta\hat B^\dag \Delta\hat B} \ge
\im^2 \mean{\Delta\hat A^\dag \Delta\hat B}
\end{equation}
and can also be obtained if the real part of
$\mean{\Delta\hat A^\dag \Delta\hat B}$
is not omitted when deriving HUR using the Cauchy-Schwartz inequality for operators.

Note that SRUR is invariant under the whole group Sp(2, $R$) of linear canonical transformations 
whereas HUR is so only under a restricted subset of Sp(2, $R$).
The group Sp(2, $R$) consists of linear transformations 
for quadrature operators $\hat x$ (position) and $\hat p$ (momentum) 
\begin{equation}
\hat X = \left[ \begin{array}{cc} \hat x \\ \hat p \end{array}\right] 
~\rightarrow ~
\hat X' =
S \left[ \begin{array}{cc} \hat x \\ \hat p \end{array}\right],
\end{equation}
which preserves the canonical commutation relation 
\begin{equation}
[\hat X_i, \hat X_j] = i\, \Omega_{i j}, \quad
\Omega = \left[ \begin{array}{cc} 0 & 1 \\ -1 & 0 \end{array}\right] .
\label{eq:ccr}
\end{equation}
In other words, $S\in {\rm Sp}(2,\,R)$ has the property of 
$S \Omega S^T = \Omega$ (or $S^T \Omega S = \Omega$).
In two dimensional case 
${\rm Sp}(2,\,R) = {\rm SL}(2,\,R)$ [${\rm SL}(n,\,F)$: $n$-dimensional special linear group over a field $F$]---hence, 
Sp(2, $R$) is a three-parameter group---though 
in general ${\rm Sp}(2n,\,R) \subset {\rm SL}(2n,\,R)$.
Each element of Sp(2, $R$) is directly related to its unitary representation $\hat{U}(S)$ as
\begin{equation}
\hat{U}^\dag(S)\, \left[ \begin{array}{cc} \hat x \\ \hat p \end{array}\right] \hat{U}(S) =
S \left[ \begin{array}{cc} \hat x \\ \hat p \end{array}\right].
\label{eq:sympxp}
\end{equation}
It also connects the density matrix with its Wigner function via Weyl-Wigner transform
\begin{equation}
\rho' = \hat{U}(S)\, \rho \hat{U}^\dag(S) 
\Longleftrightarrow
W_{\rho'} (X) = W_{\rho}(S^{-1} X)
\end{equation}
where $X=(x,p)$.

In this article,
we just sketch the formulation of generalized SRUR following the steps taken by Ivan \etal \cite{Ivan12a
}. Readers who are interested in more details including its group properties may refer to Refs. \cite{Ivan12a
}.
We may start by extending the quadrature operators $\hat x$ and $\hat p$ to their higher-order forms such as $\hat x^m \hat p^n$.
For a systematic extension, 
a certain kind of ordering should be taken into account and 
Weyl ordering of $\hat x$ and $\hat p$ renders such higher-order products hermitian.
Although the hermiticity of operators is not necessarily required,
Weyl ordered ``monomials'' transform exactly the same as their classical variables do under linear canonical transformations. So the desired UR can also be transformed in a simple manner under such transformations.
We define the Weyl-ordered \emph{monomial} observable
$\hat T_{jm}$ $\left(j=\half,1,\frac{3}{2}, \cdots, m=j,j-1,\cdots,-j\right)$ 
as symmetrized homogeneous product of $\hat x^{j+m} \hat p^{j-m}$ \cite{Ivan12a
}, that is,
\begin{eqnarray}
\left(\hat{T}_{\half m}\right)= \left[  
 \begin{array}{c} \hat{x}\\ \hat{p} \end{array} \right], \quad
\left(\hat{T}_{1 m}\right)= \left[
 \begin{array}{c} \hat{x}^2\\ 
  \half (\hat{x}\hat{p} + \hat{p}\hat{x}) \\ \hat{p}^2
 \end{array} \right], \nonumber \\
\left(\hat{T}_{\frac{3}{2} m}\right)= \left[
 \begin{array}{c} \hat{x}^3\\ 
  \frac{1}{3}(\hat{x}^2\hat{p} + \hat{x}\hat{p}\hat{x}+ \hat{p}
  \hat{x}^2)\\
  \frac{1}{3}(\hat{x}\hat{p}^2 + \hat{p}\hat{x}\hat{p}+ \hat{p}^2
  \hat{x})\\
  \hat{p}^3
 \end{array} \right], \quad \cdots.
\label{eq:tjm}
\end{eqnarray}
Note that the angular momentum notation is used for $\hat T_{jm}$
since its product law is determined by the Clebsch-Gordan (CG) coefficients of SU(2), \ie, 
\begin{eqnarray}
&\hat{\tau}_{jm}\,\hat{\tau}_{j' m'} 
= \sum _{j'' = |j-j'|}^{j+j'} \left(\frac{i}{2} \right)^{j+j'-j''}
C^{j\,j'\,j''}_{m\,m'\, m+m'}\, \hat{\tau}_{j'',\,m +m'} \nonumber \\
\times &\sqrt{\frac{(j+j' + j'' +1)!}{(2j'' +1)(j+j'-j'')! (j'+j'' -j)! (j'' +j -j')!}}, \\
&\hat{\tau}_{jm} = \hat{T}_{jm}/\sqrt{(j+m)!(j-m)!},
\end{eqnarray}
where $C^{j\,j'\,j''}_{m\,m'\,m''} = \braket{jm,j'm'}{j''m''}$ is the 
CG coefficient of SU(2).
The fact that SU(2) and Sp(2, $R$) have the same product law is due to the analytic continuation between SU(2) and Sp(2, $R$) in finite dimension.
As can be seen in the above formula, 
the product  $\hat{T}_{jm}\,\hat{T}_{j'm'}$  may not generally be hermitian,
e.g.,
\begin{eqnarray*}
\hat{T}_{\half m} \hat{T}_{1 m'}
&=& \hat{T}_{\textstyle{\frac{3}{2}}\, m+m'} 
 + \frac{i}{2}(2m-m') \hat{T}_{\half\, m+m'} \\
\hat{T}_{1 m} \hat{T}_{1\, m'} 
&=& \hat{T}_{2\, m+m'} 
 + i (m-m') \hat{T}_{1\, m+m'}\\
&&+ \frac{(-1)^m}{4} (2-\delta_{m,0})\delta_{m+m',0} \hat{T}_{00}.
\end{eqnarray*}
The real and imaginary parts are discriminated when
computing the corresponding covariance matrix, as will be clarified later.
Given the Wigner function of a state,
the mean value of $\hat{T}_{jm}$ can be easily calculated since it is already of Weyl-ordered form, \ie,
\begin{equation}
\mean{\hat{T}_{jm} }_{\rho} = \int \!\!\int dx\, dp\, W_{\rho}(x,p)
 \,x^{j+m}\,p^{j-m} .
\label{eq:mvtjm}
\end{equation}

Now we are in a position to evaluate the MM in~(\ref{eq:mf}) with (hermitian) operators
\begin{equation}
\hat{\bf f} =  
\left(\Delta\hat{T}_{\half\, \half},\,
 \Delta\hat{T}_{\half\, -\half},\, \Delta\hat{T}_{1 1},\, \Delta\hat{T}_{1 0},\,\Delta\hat{T}_{1 -1}
 ,\,\cdots \right)
\label{eq:ftjm}
\end{equation}
using formulas in Eqs. (\ref{eq:tjm})-(\ref{eq:mvtjm}).
One may use 
$\hat{\bf f}' =  
(\hat{T}_{00}, \hat{T}_{\half\, \half},\,
\hat{T}_{\half\, -\half},\, \hat{T}_{1 1},\, \hat{T}_{1 0},\, \hat{T}_{1 -1},\, \cdots)$ with $\hat{T}_{00}=\hat 1$
whereby the dimension of MM increases by one. 
Note that $M_{\hat{\bf f}'}$ can turn into $M_{\hat{\bf f}}$ easily by 
the Schur complement of its (1,1)-entry.
The corresponding MM 
\begin{equation}
M_{\hat{\bf f}} = \left(\mean{\Delta \hat{T}_{jm} \Delta \hat{T}_{j'm'}} \right)
\label{eq:mftjm}
\end{equation}
has its each entry as
\begin{eqnarray}
& M_{jm,j'm'}
 = V_{jm,j'm'} + \frac{i}{2} \, \Omega_{jm,j'm'},  
\end{eqnarray}
where
\begin{eqnarray} 
& V_{jm,j'm'} =\half 
 \mean{\{\hat{T}_{jm},\,\hat{T}_{j'm'} \}} 
 - \mean{\hat{T}_{jm}}\,\mean{\hat{T}_{j'm'}}, \\
&\Omega_{jm,j'm'} = \frac{1}{i} \mean{[\hat{T}_{jm},\,\hat{T}_{j'm'}]}.
\end{eqnarray}
Here,
$j,\,j' = \half,\,1,\,\cdots$, $m = j, \cdots, -j$, 
and $m' = j', \cdots, -j'$
and the matrix $V$ ($\Omega$) is the real symmetric (imaginary anti-symmetric) part of $M_{\hat{\bf f}}$. 
Note that for $j=1/2$, the matrix $(\Omega_{jm,j'm'})$ is nothing but $\Omega$ in ~(\ref{eq:ccr}).

Since the matrix in (\ref{eq:mftjm}) is infinite dimensional,
one should consider its finite truncated version for practical use, thereby producing a hierarchy of URs.
The first one starts with
$\hat{\bf f} = 
\left(\Delta\hat{T}_{\half\,m} \right) \equiv
\left(\Delta\hat{T}_{\half\,\half},\,\Delta\hat{T}_{\half\, -\half} \right) = \left(\Delta\hat{x},\Delta\hat{p}\right)$
and this observable set leads to the original SRUR, 
which we label as ($J=1/2$)-th covariance matrix (CM)
\begin{equation}
M_\half(\rho) 
\equiv \left[ \begin{array}{cc} 
\mean{(\Delta\hat x)^2} & \mean{\Delta\hat x \Delta\hat p} \\
\mean{\Delta\hat p \Delta\hat x} & \mean{(\Delta\hat p)^2} 
\end{array} \right].
\end{equation}
The next ($J=1$)-th CM is constructed with 
\begin{equation}
\hat{\bf f} = 
\left(\Delta\hat{T}_{\half\,m},\, \Delta\hat{T}_{1\,m}\right) \equiv
\Big(\Delta\hat{T}_{\half\,\half},\,\Delta\hat{T}_{\half\, -\half},\,
\Delta\hat{T}_{1,\,1},\,\Delta\hat{T}_{1\,0},\,,\Delta\hat{T}_{1\,-1}\Big)
\end{equation}
and reads
\begin{equation}
M_1 (\rho) \equiv
 \left[\begin{array}{c|c}
 M_\half(\rho) & M_{\half, 1}(\rho)\\
 \hline
 M_{1 , \half}(\rho) & M_{1,1}(\rho)
\end{array} \right],
\end{equation}
where
\begin{eqnarray}
M_{\half, 1}(\rho) =
M_{1,\half}^\dag(\rho) =
\left[ \begin{array}{ccc}
\mean{\Delta\hat x \Delta\hat T_{1 1}} & \mean{\Delta\hat x \Delta\hat T_{1 0}} &
\mean{\Delta\hat x \Delta\hat T_{1 -1}} \\
\mean{\Delta\hat p \Delta\hat T_{1 1}} & \mean{\Delta\hat p \Delta\hat T_{1 0}} &
\mean{\Delta\hat p \Delta\hat T_{1 -1}} \nonumber
\end{array} \right].
\end{eqnarray}
Now if we label $M_{\hat{\bf f}}$ with 
$\hat{\bf f} = 
\left(\Delta\hat{T}_{\half\,m},\,\cdots,\, \Delta\hat{T}_{J\,m} \right)\equiv
\left(\Delta\hat{T}_{\half\, \half}, \,
\cdots,\, \Delta\hat{T}_{J,-J} \right)$
as $J$-th CM $M_J$,
we can systematically extend the CM from $J$-th one to $(J+1/2)$-th one, by adding the operators $\Delta\hat{T}_{J+\half,m}$ ($m=-J-\half,\cdots,J+\half$), as
\begin{eqnarray}
M_{J + \half}  &= \left[ 
 \begin{array}{c|c} 
 M_J & 
M_{\half:J,J+\half} \\
 \hline 
 M_{J+\half,\half:J} & 
 M_{J+ \half, J+ \half}
 \end{array} \right].
\label{eq:mext}
\end{eqnarray}
Here  
$M_{J+\half,\half:J} = M^\dag_{\half:J,J+\half} =\left[ 
\begin{array}{ccc}
 M_{J+ \half, \half} 
 & \cdots & 
 M_{J + \half, J}
 \end{array} 
\right]$
is the lower left $(2J+2) \times N_{J}$ off-diagonal block matrix with $N_{J}=J(2J+3)$.
Equipped now with CM truncated up to $J$-th monomial observables,
the desired hierarchy of SRUR can be phrased in the form of $N_{J}$-dimensional matrix as
\begin{equation}
M_J (\rho) = 
V_J(\rho)+ \frac{i}{2}\,{\Omega}_J (\rho) \ge 0 .
\label{eq:srurext}
\end{equation} 
where 
$M_J = (M_{jm,j'm'})$ and $V_J$ and $\Omega_J$ are defined in the same way \cite{Ivan12a
}.

In order to check the nonnegativity of the above CM,
one may use the Sylvester criterion
\cite{Shchukin0511,Miranowicz09,Miranowicz06,Horn85},
which states that 
a hermitian matrix is nonnegative if and only if \emph{all} its 
principal minors are nonnegative.
Alternatively,
one may adopt another simpler criterion,
which states that
a hermitian matrix is nonnegative if and only if one of its leading principal submatrix is positive and the corresponding Schur complement is nonnegative.
In our case,
this can be formulated as follows.
After checking the positivity of $M_J$,
one can proceed to check the nonnegativity of its Schur complement in the whole matrix of $M_{J + \half}$ in (\ref{eq:mext}), namely,
\begin{equation}
M_{J+\half \vert J} \equiv 
M_{J+\half, J+\half} - M_{J+\half,\half:J} \, M_J^{-1}\,M_{J+\half,\half:J}^{\dagger} \ge 0.
\label{eq:sc}
\end{equation}
Here $M_J$ is the $N_J \times N_J$ leading principal submatrix 
($J=\half,\,1,\,\frac{3}{2},\,\cdots$).
In case that $M_J$ has a zero eigenvalue(s), 
one can simply split it into its null space and the remaining invertible one 
and applying the above inequality by ignoring the null-space-related block matrices.
Or equivalently, and more simply, one can replace $M_J^{-1}$ in the above by the Moore-Penrose inverse of $M_J$ \cite{Horn85
}.

Thus far,
we have sketched the procedure to construct CM and to check its positivity in a systematic manner. The next part is devoted to mentioning its covariance property.
We begin by noting that 
the span of $\hat{T}_{jm}$ (of the same $j$) is invariant under the unitary operator $\hat{U}(S)$ and 
that each transformed element is in that span, \ie,
\begin{eqnarray}
\hat{U}^\dag(S)\, \hat{T}_{jm} \hat{U}(S) =
 \sum_{m'= -j}^{j} K^{(j)}_{m m'} (S)\,\hat{T}_{jm'}.
\end{eqnarray}
Comparing this with (\ref{eq:sympxp}),
we notice that $K^{(1/2)}(S)=S$.
This $(2j+1)$-dimensional $K^{(j)}(S)$ is the real (nonunitary) irreducible representation of Sp(2,\,$R$) for $\hat{T}_{jm}$ and can be obtained by the same transformation rule of classical \emph{monomial}.
For example, for arbitrary $S\in {\rm Sp}(2,\,R)$,
$K^{(1)}(S)$ can be obtained by the transformation rule of the
monomial set of $(x^2,xp,p^2)$,
\begin{equation}
\left[\begin{array}{c} x'\\ p' \end{array} \right]
= S \left[\begin{array}{c} x\\ p \end{array} \right]
~\longrightarrow~
\left[\begin{array}{c} x'^2\\ x'p'\\ p'^2 \end{array} \right]
= K^{(1)}(S) \left[\begin{array}{c} x^2\\ xp\\ p^2 \end{array} \right],
\end{equation}
which implies that
\begin{equation}
S = \left[\begin{array}{cc} a & b \\ c & d \end{array} \right]
~\longrightarrow~
K^{(1)}(S) = \left[\begin{array}{ccc} 
a^2 & 2ab & b^2 \\ 
ac & ad+bc & bd \\
c^2 & 2cd & d^2 \end{array} \right]
\end{equation}
and $K^{(J)}(S)$ with higher $J$ can be derived in a similar way.

Owing to the above covariance property,
the resulting transformation rule of $M_J$ becomes simple enough,
that is,
\begin{equation}
\rho\rightarrow \hat U (S)\,\rho\,\hat U(S)^\dag 
\Rightarrow 
M_J \rightarrow K_J(S) M_J K_J^T(S)
\end{equation}
where 
\begin{equation}
K_J(S)= K^{(\half)}(S)\oplus K^{(1)}(S)\oplus \cdots \oplus K^{(J)}(S).
\label{eq:kj}
\end{equation}
For $J=1$ case, as an illustration,
the relevant block matrices transform respectively as
\begin{equation}  
M_\half \rightarrow S M_\half S^{T}, \quad
M_1 \rightarrow K^{(1)} M_1 K^{(1)T}, 
M_{1,\half} \rightarrow K^{(1)} M_{1,\half} S^{T}. \nonumber
\end{equation}

\section{Inseparability criterion using single-mode Schr\"odinger-Robertson uncertainty relation}
\label{sec:ec1}

It is well known that 
partial transposition can map a bipartite inseparable state to a form not admissible as a legitimate quantum state.
That is, partially transposed (PT) density matrix 
$\rho^\Gamma$ of a bipartite entangled state $\rho$ can possess a negative eigenvalue 
and this negativity is also passed on to its MM, $M_{\hat {\bf f}} (\rho^\Gamma)$.
Formally, its negativity is a
necessary and sufficient condition for NPT \cite{Shchukin0511,Miranowicz09}.
In our framework, in order to detect the entanglement of $\rho$,
we must show the condition $M_J (\rho^\Gamma) < 0$
---or equivalently find its negative eigenvalue(s)---at a certain level of $J$.
However, as addressed in the next section,
testing the positivity of two-mode SRUR is rather demanding.
For instance, 
in order for $M_J (\rho^\Gamma)$ to be probed up to the fourth-order ($J=1$),
one should search the eigenvalues of a fourteen---or 
ten if one considers the Schur complement of $M_{\half} (\rho^\Gamma)$
[see (\ref{eq:sc})]---dimensional matrix (this will be clarified in the next section), which is practically burdensome.

A possible approach to bypass this issue is to use a single-mode marginal distribution of two-mode $\rho^\Gamma$.
One first changes quadrature variables to a new set by
\begin{equation}
x_\pm = \frac{x'_1 \pm  x'_2}{\sqrt{2}}, \quad
p_\pm = \frac{p'_1 \pm p'_2}{\sqrt{2}} ,
\label{eq:changevariables}
\end{equation}
where 
$x'_j=\cos\theta_j x_j + \sin\theta_j p_j$, 
$p'_j =\cos\theta_j p_j - \sin\theta_j x_j$.
By changing the arguments through the above,
one can get a new Wigner function $W(x_+,p_+,x_-,p_-)$ from the original Wigner function $W(x_1,p_1,x_2,p_2)$ of $\rho$, which essentially corresponds to a beam-splitting operation. With the commutation relation $[x_\pm, p_\pm]=i$,
$W(x_+,p_+,x_-,p_-)$ can be considered the Wigner function associated with the two-mode quadratures $(x_+,p_+)$ and $(x_-,p_-)$.
If one applies to it partial transposition \cite{Simon00,Walborn09}
by changing the sign of $p_2'$ in (\ref{eq:changevariables}), 
then
\begin{equation}
W(x_+,p_+,x_-,p_-) ~\rightarrow~ W(x_+,p_-,x_-,p_+). 
\end{equation}
If the original Wigner function $W(x_1,p_1,x_2,p_2)$ describes a separable state, the resulting $W(x_+,p_-,x_-,p_+)$ must also be a \emph{bona fide} Wigner function. Therefore, its marginal single-mode distributions 
\begin{equation}
W(x_\pm,p_\mp) = \int\!\!\int dx_\mp dp_\pm W(x_+,p_-,x_-,p_+)
\end{equation}
are also legitimate Wigner functions.
From this we can compute $M_{\hat {\mathbf f}}(\rho^{M\Gamma})$
(hereafter, we denote $\rho^{M\Gamma}$ as the marginal PT density matrix)
and check its nonnegativity as formulated in the previous section, \ie,
\begin{equation}
M_J (\rho^{M\Gamma})
= V_J (\rho^{M\Gamma})+ \frac{i}{2}\, \Omega_J(\rho^{M\Gamma}) \ge 0.
\label{eq:ecsrur}
\end{equation}

\begin{figure}[t]{
\includegraphics[width=\columnwidth]{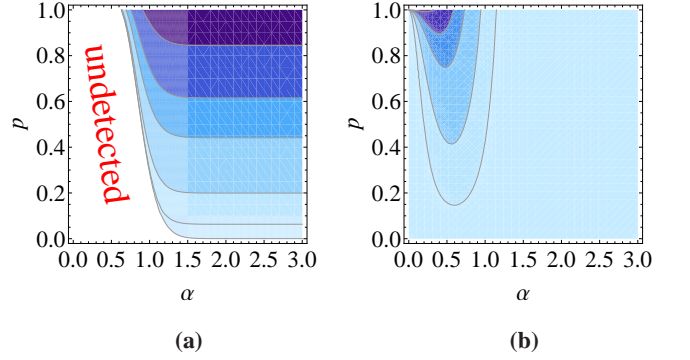}}
\leftline{\hskip2.3cm {\bf (a)} \hskip4.cm {\bf(b)} }
\caption{(Color online)
(a) Sum of entropies subtracted by the entropic bound, \ie, $H[P(x_-)] + H[P(p_+)] - \ln (\pi e)$ as a function of 
the amplitude $\alpha$ and the coherence parameter $p$ of the dephased cat
$\rho_\mathrm{cat}$.
(b) Determinant of the fourth-order MM of marginal PT dephased cat, \ie,
$\det [M_{1|\half}(\rho_\mathrm{cat}^{M\Gamma})]$
using the same observables $x_-$ and $p_+$ as in (a).
In each panel, the contours are
--\,0.2, --\,0.1, --\,0.05, --\,0.01, --\,0.001, and 0 
from above [note that 0 is missing in (b)] and
a negative value implies that the respective criterion detects entanglement.
Notice that the entropic criterion cannot detect entanglement for small $\alpha$ 
while the fourth-order MM criterion does for all $\alpha$ and $p$.
}
\label{fig:cat}
\end{figure}

Let us now apply the above SRUR to a dephased cat state
\begin{align}
\rho_\mathrm{cat} = \, &\mathcal{N} \big[ \ket{\alpha,\alpha}\bra{\alpha,\alpha} + \ket{-\alpha,-\alpha}\bra{-\alpha,-\alpha} \nonumber \\
&-  p (\ket{\alpha,\alpha}\bra{-\alpha,-\alpha} +\ket{-\alpha,-\alpha}\bra{\alpha,\alpha} ) \big],
\label{eq:cat}
\end{align}
where the amplitude $\alpha$ is assumed to be real, $0\le p \le 1$ represents the degree of coherence,
and $\mathcal{N} = 1/[2-2p \exp(-4\alpha^2)]$ is the normalization factor.
Note that $\rho_\mathrm{cat}$ is separable only when $p=0$ and  
its inseparability is not detected by the second-order criteria and that
in Ref. \cite{Walborn09} an entropic UR criterion is introduced aiming at its detection. That is, 
\begin{equation}
H[P(x_\pm)] + H[P(p_\mp)] \ge \ln (\pi e)
\label{eq:eur}
\end{equation}
where 
\begin{equation}
H[P(q)] = - \int dq P(q) \ln P(q)
\end{equation}
is the Shannon entropy for a probability distribution $P(q)$.
However, even this entropic criterion 
detects its inseparability only for large $\alpha$ and $p$ [see Fig. \ref{fig:cat}(a)] \cite{Walborn09}.
We now demonstrate that the fourth-order SRUR formulated here can detect
the entanglement of $\rho_\mathrm{cat}$ for any values of $\alpha$ and $p$.
As aforementioned,
$M_{\half}(\rho^{M\Gamma}_{\mathrm{cat}}) > 0$, 
which can be seen from its positive determinant 
$\det (M_{\half}) = 2 \alpha^2 (2\mathcal{N}-1) >0$.
Hence, we proceed to check the next hierarchy, \ie, $J=1$ case
\begin{equation}
M_{1 \vert \half} = 
M_{1,1} - M_{1,\half} M_{\half}^{-1} M_{\half,1} \ge 0.
\label{eq:srur4th}
\end{equation}
If we choose the observables $x_-$, $p_+$ and the parameters $\theta_1 = \theta_2 = 0$ in (\ref{eq:changevariables}),
\ie, $x_- = (x_1-x_2)/\sqrt{2}$ and $p_+ = (p_1+p_2)/\sqrt{2}$, 
we get $M_{1,\half} = 0$ and hence 
\begin{equation*}
M_{1 \vert \half} = M_{1,1} = \half
\left[\begin{array}{ccc} 
1 & i & -1 \\ 
-i & 1+2d & i (1+4d)\\
-1 & -i (1+4d) & 1+8d(1-4\mathcal{N}\alpha^2) \end{array} \right], \nonumber
\end{equation*}
where 
$d=\det (M_{\half})$.
This matrix has two positive and one negative eigenvalues and
therefore is always negative for all $\alpha$ and $p$.
In Fig. \ref{fig:cat},
we show both the sum of entropic uncertainties in (\ref{eq:eur}) and 
the determinant $\det (M_{1|\half}) = -8 \mathcal{N} \alpha^2 d^2$.
It is not necessary here but 
the negativity of (\ref{eq:srur4th}) may also be optimized by introducing another local transformations. 
Since the local symplectic group Sp(2,\,$R$) $\otimes$ Sp(2,\,$R$), a subgroup of the full symplectic group Sp(4,\,$R$),
has six parameters,
one can use another four parameters, \ie, 
another two local rotations along with two local squeezing actions 
after the two rotations related to $\theta_1$ and $\theta_2$ \cite{Serafini07}.

As can be seen from the figure,
the fourth-order SRUR fully detects the inseparability of $\rho_\mathrm{cat}$
whereas the entropic criterion does not.
In view of this, it is worth emphasizing that 
although an entropic UR also involves higher-order moments of correlations in a certain form,
it does not fully reflect the specific-order---fourth-order
in this case---moments.
This observation seems to be also supported by comparing 
the undetected region (the white region) in Fig. \ref{fig:cat}(a) and
the region of relatively large negativity of $M_{1 \vert \half}$ in Fig. \ref{fig:cat}(b) .
We suppose that the entanglement of a dephased cat is coded relatively more in
the fourth-order correlation when its size is not so large.

As another advantage over the entropic criterion,
our method may provide an analytical result in algebraically simpler form.
Even when viewed from the perspective of numerical cost,
since the dimension of $M_{1 \vert \half}$ is as small as three,
finding its eigenvalues does not require much effort compared to calculating the entropies.
Moreover, let us remark further on a possible practical efficiency.
According to Sylvester criterion mentioned in the previous section,
we can search the negativity of $M_J (\rho^{M\Gamma})$ 
by choosing a certain partial number of rows (columns) and need not investigate the whole matrix.
Indeed, in the above case,
the principal submatrix of $M_{1, 1}$ built by choosing the second and third rows/columns suffices to detect inseparability.

Finally, we mention that
a dephased cat may be classified as an inseparable state whose entanglement is revealed by a fourth-order-quadrature-moment criterion but not by any lower-order one.
In this sense, we might call a dephased cat as a \emph{fourth-order entangled} state.

\section{Generalized Schr\"odinger-Robertson uncertainty relation for two-mode case 
and the corresponding inseparability criterion}
\label{sec:srur2}

In the previous sections,
we have addressed the single-mode SRUR and used it for an inseparability criterion
by applying it to a marginal PT state and checking its negativity.
Despite its better performance over the entropic criterion, however, 
there can be states which are not detected by the criterion 
yet whose entanglement is still coded in the fourth-order correlation.
Indeed, such states exist and will be introduced later.
Thus, to conclusively determine whether or not
the entanglement is due to a specific order correlation, 
we should have a full two-mode correlation criterion without marginalization.

With this motivation,
we extend the single-mode SRUR derived by Ivan \etal in Sec. \ref{sec:srur1}
to a two-mode case to be employed as an inseparability criterion.
Apparently, it seems straightforward as we only need to include
every observable of two modes up to the desired order.
For instance, if the fourth order SRUR is to be addressed, 
it is necessary to construct the operator set $\hat {\bf f}$ 
by all observables up to the second-order, namely, 
$\hat T^{A}_{\half m},\,\hat T^{B}_{\half m} ,\,\hat T^{A}_{\half m} \hat T^{B}_{\half m'},\,\hat T^{A}_{1 m},\,\hat T^{B}_{1 m}$,
where $A$ and $B$ denote two distinct modes.

As one may readily appreciate, for a systematic extension,
a specific ordering of observable operators is needed
for an efficient construction of MM and more importantly for symplectic covariance.
We explicitly give an ordering for generic observables
$\{\hat T^{A}_{j_A m_A}\hat T^{B}_{j_B m_B}\}$ ($\hat T^{A(B)}_{00}= \hat 1$ as before).
$\hat T^{A}_{j_A m_A}\hat T^{B}_{j_B m_B}$ comes before
$\hat T^{A}_{j'_A m'_A}\hat T^{B}_{j'_B m'_B}$ 
if and only if the first non-zero difference of
$j'_A+j'_B - (j_A+j_B),~j_A - j'_A,~ m_A - m'_A,~m_B - m'_B$ is positive.
In the case of the observables for $M_{J=1}$, 
for example, we have the following ordering
\begin{eqnarray}
&\hat T^{A}_{\half \half},\,\hat T^{A}_{\half -\half},\,
\hat T^{B}_{\half \half},\,\hat T^{B}_{\half -\half},\,
\hat T^{A}_{11},\,\hat T^{A}_{10},\,\hat T^{A}_{1-1},\,
\hat T^{A}_{\half \half}\hat T^{B}_{\half \half},\,\nonumber\\
&\hat T^{A}_{\half \half}\hat T^{B}_{\half -\half},\,
\hat T^{A}_{\half -\half}\hat T^{B}_{\half \half},\,
\hat T^{A}_{\half -\half}\hat T^{B}_{\half -\half},\,
\hat T^{B}_{11},\,\hat T^{B}_{10},\,\hat T^{B}_{1-1},
\label{eq:tjmab}
\end{eqnarray}
where ``$\Delta$'' notations are omitted for brevity.
The remaining procedure of obtaining the two-mode CM $M_J(\rho)$ and checking its covariance property under symplectic transformations are straightforward
as before.
One can easily get the two-mode version of $K_J(S)$ in (\ref{eq:kj})
for the full symplectic group Sp(4, $R$).

Equipped with the full two-mode SRUR, 
the construction of the corresponding inseparability criterion is also straightforward.
We have only to check the positivity of the two-mode CM $M_J(\rho^\Gamma)$ where $\rho^\Gamma$ is the PT density matrix of $\rho$
(this time a full two-mode one, not a marginal one).
It is also necessary to consider the symplectic transformations 
which leave invariant $M_J(\rho^\Gamma)$ of a separable state $\rho$.
One can readily see that under Sp(2,\,$R$) $\otimes$ Sp(2,\,$R$), 
the aforementioned $K_J(S)$ reduces to 
\begin{eqnarray}
K_J(S) &=& K_A^{(\half)}(S)\oplus K_B^{(\half)}(S)\oplus K_A^{(1)}(S) \oplus
\left[ K_A^{(\half)}(S)\otimes K_B^{(\half)}(S) \right] \nonumber \\
&\oplus&  K_B^{(1)}(S) \oplus \cdots \oplus K_B^{(J)}(S).
\label{eq:kj2}
\end{eqnarray}

Unfortunately, however, 
inseparability criterion using this two-mode SRUR might be considered not practically useful 
since the size of relevant CM grows huge with dimension as $J$ increases.
In more detail, 
in order to attain 4$J$-th-order full SRUR in this hierarchy,
one must compute 
$\frac{1}{6} J(2J+5)(2J^2+5J+5)$ dimensional square matrix---\eg,
$34\times34$ matrix for the case of the sixth-order SRUR ($J=3/2$)---which is very demanding.
Even worse, the dimension grows as $\sim J^4$ 
while that of the marginal one just as $\sim J^2$.
One resolution to this issue is to use a principal submatrix,
whose usage is justified in the previous section.

\begin{figure}[t]{
\includegraphics[width=\columnwidth]{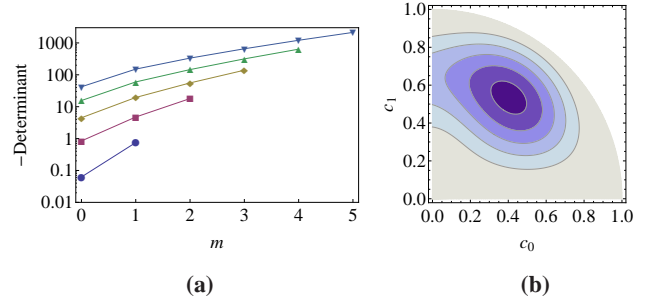}}
\leftline{\hskip2.3cm {\bf (a)} \hskip4.cm {\bf(b)} }
\caption{(Color online)
(a) Negative determinant of the submatrix of the fourth-order MM of PT beam-split number state $\hat B \ket{n,m}$. 
From below, the lines denote the case of $n=$ 1, 2, 3, 4, 5.
(b) Determinant of the the submatrix of fourth-order MM of PT number-correlated state
$c_0 \ket{0,0} + c_1 \ket{1,1} + c_2 \ket{2,2}$.
The contours are --10, --20, --30, --40, --50; 
the darker the region is, the large its absolute value is.
For both cases, 
a negative value indicates that the entanglement of the corresponding state is detected.
}
\label{fig:bnsncs}
\end{figure}

As mentioned in the early part of this section,
there exist
certain entangled states that are not detected by the marginal fourth-order SRUR criterion 
but detected by the full two-mode version.
We illustrate this by introducing
beam-split number states (BNSs) $\hat B \ket{n,m}$ 
($\hat B$: the unitary operator of a beam splitter, 
$\ket{n,m}$: a two-mode number state).
This class of states is scarcely detected by (\ref{eq:srur4th})
but detected by its full two-mode version.
Furthermore, as mentioned in the previous paragraph,
the inseparability of BNSs can be practically detected by its submatrix:
the four observables 
$\hat T^{A}_{11},\,\hat T^{A}_{\half \half}\hat T^{B}_{\half \half},\,
\hat T^{A}_{\half \half}\hat T^{B}_{\half -\half},\,
\hat T^{A}_{\half -\half}\hat T^{B}_{\half -\half}$ 
in (\ref{eq:tjmab}) are sufficient to build a CM that detects the inseparability for any photon numbers $n$ and $m$.
In more detail, the corresponding submatrix of $M_{1,1}$ has 
three positive and one negative eigenvalues 
and hence its negativity can again be checked by its determinant
[See Fig. \ref{fig:bnsncs}(a)].
Interestingly, in order to detect it,
the author in Ref. \cite{Namiki12} has resorted to a
fourth-order HUR  in the case of $n \ne m$ 
using the operators from su(2) and su(1,1) algebras \cite{Agarwal05,Hillery06,Nha06}
while, in the case of $n=m$, to an eighth-order HUR
using their higher-order extended operators \cite{Sun09}.
However, similarly as in the case of a dephased cat, we stress that
the BNSs may be categorized as \emph{fourth-order entangled} states.

In order to further illustrate the power of our two-mode approach, 
we here present another example of the fourth-order entangled states 
that are not detected by the marginal fourth-order criterion but by the full two-mode one. 
We consider a class of photon-number entangled state (PNES), i.e. 
$\ket{\Psi}=\Sigma_n c_n \ket{n,n}$. The class of PNES states has drawn much attention as it constitutes an important resource for continuous-variable quantum communication including both Gaussian (two-mode squeezed state) and non-Gaussian entangled states \cite{Lee12}. As an example, a truncated PNES
$\ket{\Psi_2}=c_0 \ket{0,0} + c_1 \ket{1,1} + c_2 \ket{2,2}$ is here investigated where
$c_0^2 + c_1^2 + c_2^2 = 1$.
Like the case of BNSs, 
the inseparability can be detected by its submatrix; 
however, for this case, by using five observables 
$\hat T^{A}_{11},\,\hat T^{A}_{1-1},\,
\hat T^{A}_{\half \half}\hat T^{B}_{\half -\half}\,
\left(\textrm{or } \hat T^{A}_{\half -\half}\hat T^{B}_{\half \half} \right),\,
\hat T^{B}_{11},\,\hat T^{B}_{1-1}$.
Since the corresponding submatrix has 
four positive and one negative eigenvalues,
its negativity can also be checked by its determinant. 
As can be seen in Fig. \ref{fig:bnsncs}(b), 
the entanglement of the class of PNES $\ket{\Psi_2}$ is again fully detected by our criterion.

\section{Comparison with other criteria}
\label{sec:comparison}

Let us now compare our formalisms with other existing second-
as well as higher-order criteria 
in a unified view of inseparability criterion based on correlation moments.
Shchukin and Vogel (SV) proposed a hierarchy of inseparability criteria using moments of annihilation and creation operators \cite{Shchukin0511}.
They construct a matrix of such moments of a PT density matrix $\rho^\Gamma$ 
from an original inseparable state $\rho$ and 
note that a (appropriately truncated) leading principal minor of the (infinite-dimensional) matrix of moments can reveal negativity 
and hence the inseparability.
According to SV criterion,
Simon's criterion \cite{Simon00}
\begin{equation}
\mean{(\Delta \hat{X})^2} + \mean{(\Delta \hat{X}')^2} \ge
|c_1 d_2 - c_2 d_1| + |c_3 d_4 - c_4 d_3|
\end{equation}
for all real $c_i$'s and $d_i$'s,
where
$\hat X = c_1\hat x_1 + c_2\hat p_1 + c_3\hat x_2 + c_4\hat p_2$ and
$\hat X' = d_1\hat x_1 + d_2\hat p_1 + d_3\hat x_2 + d_4\hat p_2$,
is equivalent to 
\begin{equation}
M_{\hat {\bf f}} (\rho^\Gamma)\ge 0
\label{eq:ecmm}
\end{equation}
with 
$\hat{\bf f} = (\Delta\hat a,\Delta\hat a^\dagger,\Delta\hat b,\Delta\hat b^\dagger)$.
Likewise, SV showed that Duan's criterion \cite{Duan00}
\begin{equation}
\mean{(\Delta \hat{X})^2} + \mean{(\Delta \hat{X}')^2} \ge c^2 + c^{-2},
\end{equation}
with
$\hat X = c \hat x_1 + c^{-1} \hat x_2$ and
$\hat X' = c \hat p_1 - c^{-1} \hat p_2$
is, by the optimized value of 
$c^2 = \sqrt{\mean{\Delta\hat b^\dag \Delta\hat b} / \mean{\Delta\hat a^\dag \Delta\hat a}}$,
equivalent to 
\begin{equation}
\mean{\Delta\hat a^\dag \Delta\hat a} \mean{\Delta\hat b^\dag \Delta\hat b} \ge
\re^2 \mean{\Delta\hat a \Delta\hat b},
\end{equation}
and can be refined to equation (\ref{eq:ecmm}) with 
$\hat {\bf f} = (\Delta\hat a,\Delta\hat b)$.

In light of SV criterion, 
$J=1/2$ case of equation (\ref{eq:ecsrur}) is equivalent to Simon's criterion 
if probed by varying $\theta_1$ and $\theta_2$ together with additional local squeezings.
When it comes to the fourth-order criterion in SV framework,
one should include apart from 
$\hat a,\hat a^\dag,\hat b,\hat b^\dag$ (``$\Delta$'' is omitted as before)
additional ten operators into $\hat {\bf f}$ in (\ref{eq:ecmm}) for full inspection, namely,
\begin{equation}
\hat a^2,\,\hat a^\dag \hat a,\,\hat a^{\dag 2},\,
\hat a \hat b,\,\hat a^\dag \hat b,\,\hat b^2,\,
\hat a \hat b^\dag,\,\hat a^\dag \hat b^\dag,\,
\hat b^\dag \hat b,\,\hat b^{\dag 2}.
\end{equation}
Note that the number of these necessary operators is equivalent to that of (\ref{eq:tjmab}) and hence 
the two-mode fourth-order SRUR is equivalent to SV criterion employing 14 operators.
In fact, 
they are equivalent up to any order 
since the number of independent quadrature operators 
$\hat x^A, \hat p^A, \hat x^B, \hat p^B$ is the same as that of 
the (independent) annihilation and creation operators 
$\hat a, \hat a^\dag, \hat b, \hat b^\dag$
and the whole possible combinations of those operators are considered in the MM.
However,
it is worth stressing that
the SRUR criterion is based on \emph{hermitian} operators 
and
transforms intuitively under local symplectic group Sp(2,\,$R$) $\otimes$ Sp(2,\,$R$).

Even though the case of $J=1$ in (\ref{eq:ecsrur}) or equivalently (\ref{eq:srur4th})
is obviously weaker than the above two full fourth-order criteria,
it is arguably said to be more practical in terms of computational cost.
For other fourth-order or even higher-order criteria, 
interested readers may refer to \cite{Miranowicz10}, wherein
inseparability criteria are analyzed not only in terms of SV criterion 
but also in terms of SV \emph{nonclassicality} criterion, 
which is based on a MM of normally ordered operators.

\section{Summary}
\label{sec:conclusion}
In this paper we have presented an inseparability criterion based on the recently derived generalized Schr\"odinger-Robertson uncertainty relation (SRUR) \cite{Ivan12a
}.
This generalized SRUR that involves two orthogonal quadratures to arbitrary high orders has a hierarchy naturally containing the original (second-order) SRUR as the lowest one.
Employing the single-mode SRUR,
we have first proposed a hierarchy of inseparability criterion using a marginal single-mode distribution of partially transposed state. This turns out to successfully detect the entanglement of a certain non-Gaussian continuous-variable (CV) state that is not fully detected by other second-order and entropic criteria.
In particular, the entropic criterion also addresses higher-order correlations of a CV state in a specific form, however, the above example amounts to illustrating that 
the entropic criterion does not fully reveal step by step the specific-order moments of correlation wherein entanglement is coded.

To delve into the above issue more clearly,
we have extended the single-mode SRUR to a two-mode one 
by introducing a systematic ordering for observable operators.
Based on this two-mode SRUR,
we have proposed an inseparability criterion which
can fully detect the entanglement coded in a specific order of quadrature-variable correlations.
We have also noted that
the inseparability criterion based on two-mode generalization of SRUR is equivalent to Shchukin and Vogel's criterion. 
They are both unified criteria based on moments together with the negativity of partial transposition.
In principle, SV criterion may also employ not only non-Hermitian but also Hermitian operators. 
Our formulation directly employs Hermitian operators, a hierarchy of position and momentum operators, and its full sympletic invariance is more manifest.

It may be an interesting question whether there can be a class of inseparable states 
whose entanglement is detected by a sixth-order-moment criterion 
but not by any lower one, namely, so called \emph{sixth-order entangled} states---or, 
in general, $4J(>6)$-th-order entangled states.
In this respect, we hope that our study could shed some light on unveiling the structure of entanglement and coming up with its useful classification.
Furthermore, we expect that it would also be possible to classify 
by the same reasoning, \ie, in terms of moments, or a similar one
nonclassical correlations such as nonlocality, steering and discord.

\acknowledgments
C.-W. L and H. N. were supported by NPRP Grant No. 4-520-1-083 from Qatar National Research Fund.
J.R. is supported by the Foundation for Polish Science TEAM project cofinanced by the EU European Regional Development Fund and a NCBiR-CHIST-ERA Project QUASAR.
J.B. acknowledges the financial support of the National Research Foundation of Korea (NRF) grant funded by the Korea government (MEST) (No. 3348-20100018 and No. 2010-0018295). 


\begin{thebibliography}{23}
\newcommand{\enquote}[1]{``#1''}

\bibitem{Simon00}
R.~Simon, \enquote{Peres-horodecki separability criterion for continuous
  variable systems,} Phys. Rev. Lett. \textbf{84}, 2726--2729 (2000).

\bibitem{Duan00}
L.-M. Duan, G.~Giedke, J.~I. Cirac, and P.~Zoller, \enquote{Inseparability
  criterion for continuous variable systems,} Phys. Rev. Lett. \textbf{84},
  2722--2725 (2000).

\bibitem{Mancini02}
S.~Mancini, V.~Giovannetti, D.~Vitali, and P.~Tombesi, \enquote{Entangling
  macroscopic oscillators exploiting radiation pressure,} Phys. Rev. Lett.
  \textbf{88}, 120401 (2002).

\bibitem{Raymer03}
M.~G. Raymer, A.~C. Funk, B.~C. Sanders, and H.~de~Guise, \enquote{Separability
  criterion for separate quantum systems,} Phys. Rev. A \textbf{67}, 052104
  (2003).

\bibitem{Agarwal05}
G.~S. Agarwal and A.~Biswas, \enquote{Inseparability inequalities for higher
  order moments for bipartite systems,} New Journal of Physics \textbf{7}, 211
  (2005).

\bibitem{Shchukin0511}
E.~Shchukin and W.~Vogel, \enquote{Inseparability criteria for continuous
  bipartite quantum states,} Phys. Rev. Lett. \textbf{95}, 230502 (2005).

\bibitem{Hillery06}
M.~Hillery and M.~S. Zubairy, \enquote{Entanglement conditions for two-mode
  states,} Phys. Rev. Lett. \textbf{96}, 050503 (2006).

\bibitem{Nha06}
H.~Nha and J.~Kim, \enquote{Entanglement criteria via the uncertainty relations
  in su(2) and su(1,1) algebras: Detection of non-gaussian entangled states,}
  Phys. Rev. A \textbf{74}, 012317 (2006).

\bibitem{Miranowicz09}
A.~Miranowicz, M.~Piani, P.~Horodecki, and R.~Horodecki,
  \enquote{Inseparability criteria based on matrices of moments,} Phys. Rev. A
  \textbf{80}, 052303 (2009).

\bibitem{Miranowicz10}
A.~Miranowicz, M.~Bartkowiak, X.~Wang, Y.-x. Liu, and F.~Nori, \enquote{Testing
  nonclassicality in multimode fields: A unified derivation of classical
  inequalities,} Phys. Rev. A \textbf{82}, 013824 (2010).

\bibitem{Walborn09}
S.~P. Walborn, B.~G. Taketani, A.~Salles, F.~Toscano, and R.~L. de~Matos~Filho,
  \enquote{Entropic entanglement criteria for continuous variables,} Phys. Rev.
  Lett. \textbf{103}, 160505 (2009).

\bibitem{Nha12}
H.~Nha, S.-Y. Lee, S.-W. Ji, and M.~S. Kim, \enquote{Efficient entanglement
  criteria beyond gaussian limits using gaussian measurements,} Phys. Rev.
  Lett. \textbf{108}, 030503 (2012).

\bibitem{Shchukin0501}
E.~Shchukin, T.~Richter, and W.~Vogel, \enquote{Nonclassicality criteria in
  terms of moments,} Phys. Rev. A \textbf{71}, 011802 (2005).

\bibitem{Ivan12a}
J.~S. Ivan, N.~Mukunda, and R.~Simon, \enquote{Moments of non-gaussian wigner
  distributions and a generalized uncertainty principle: I. the single-mode
  case,} J. of Phys. A: Math. Theor. \textbf{45}, 195305 (2012).

\bibitem{Shchukin0510}
E.~V. Shchukin and W.~Vogel, \enquote{Nonclassical moments and their
  measurement,} Phys. Rev. A \textbf{72}, 043808 (2005).

\bibitem{Menzel10}
E.~P. Menzel, F.~Deppe, M.~Mariantoni, M.~A. Araque~Caballero, A.~Baust,
  T.~Niemczyk, E.~Hoffmann, A.~Marx, E.~Solano, and R.~Gross,
  \enquote{Dual-path state reconstruction scheme for propagating quantum
  microwaves and detector noise tomography,} Phys. Rev. Lett. \textbf{105},
  100401 (2010).

\bibitem{Eichler11}
C.~Eichler, D.~Bozyigit, C.~Lang, L.~Steffen, J.~Fink, and A.~Wallraff,
  \enquote{Experimental state tomography of itinerant single microwave
  photons,} Phys. Rev. Lett. \textbf{106}, 220503 (2011).

\bibitem{Miranowicz06}
A.~Miranowicz and M.~Piani, \enquote{Comment on ``inseparability criteria for
  continuous bipartite quantum states'',} Phys. Rev. Lett. \textbf{97}, 058901
  (2006).

\bibitem{Horn85}
R.~A. Horn and C.~R. Johnson, \emph{Matrix Analysis} (Cambridge University
  Press, 1985).

\bibitem{Serafini07}
A.~Serafini and G.~Adesso, \enquote{Standard forms and entanglement engineering
  of multimode gaussian states under local operations,} Journal of Physics A:
  Mathematical and Theoretical \textbf{40}, 8041 (2007).

\bibitem{Namiki12}
R.~Namiki, \enquote{Photonic families of non-gaussian entangled states and
  entanglement criteria for continuous-variable systems,} Phys. Rev. A
  \textbf{85}, 062307 (2012).

\bibitem{Sun09}
Q.~Sun, H.~Nha, and M.~S. Zubairy, \enquote{Entanglement criteria and
  nonlocality for multimode continuous-variable systems,} Phys. Rev. A
  \textbf{80}, 020101 (2009).

\bibitem{Lee12}
S.-Y. Lee, J.~Park, H.-W. Lee, and H.~Nha, \enquote{Generating arbitrary
  photon-number entangled states for continuous-variable quantum informatics,}
  Opt. Express \textbf{20}, 14221--14233 (2012).

\end{thebibliography}


\end{document}